\input{aipcheck}

\documentclass[
    ,final            
  ]
  {aipproc}
\usepackage{amsmath}
\layoutstyle{6x9}

\begin{document}

\title{Wave functions for dynamically generated resonances; the two $\Lambda(1405)$ and $\Lambda(1670)$}


\author{J. Yamagata-Sekihara}{
  address={Departamento de Fisica Teorica, Universidad de Valencia}
  ,altaddress={Instituto de F{\'\i}sica Corpuscular (centro mixto CSIC-UV)\\
Institutos de Investigaci\'on de Paterna, Aptdo. 22085, 46071, Valencia, Spain }
}

\author{J. Nieves}{
  address={Instituto de F{\'\i}sica Corpuscular (centro mixto CSIC-UV)\\
Institutos de Investigaci\'on de Paterna, Aptdo. 22085, 46071, Valencia, Spain }
}

\author{E. Oset}{
  address={Departamento de Fisica Teorica, Universidad de Valencia}
  ,altaddress={Instituto de F{\'\i}sica Corpuscular (centro mixto CSIC-UV)\\
Institutos de Investigaci\'on de Paterna, Aptdo. 22085, 46071, Valencia, Spain }
}

\begin{abstract}
 In this work we develop a formalism to evaluate wave functions in momentum and coordinate space for the resonant states dynamically generated in a unitary coupled channel approach. The on shell approach for the scattering matrix, commonly used, is also obtained in Quantum Mechanics with a separable potential, which allows one to write wave functions in a trivial way. We develop useful relationships among the couplings of the dynamically generated resonances to the different channels and the wave functions at the origin. The formalism provides an intuitive picture of the resonances in the coupled channel approach, as bound states of one bound channel, which decays into open ones. It also provides an insight and practical rules for evaluating couplings of the resonances to external sources and how to deal with final state interaction in production processes. As an application of the formalism we evaluate the wave functions of the two $\Lambda(1405)$ states in the $\pi \Sigma$, $\bar{K} N$ and other coupled channels. 
It also offers a practical way to study three body systems when two of them cluster into a resonance.
\end{abstract}

\maketitle


\section{Introduction}

The chiral unitary approach to hadron dynamics has brought a new perspective to deal with the interaction of hadrons and the nature of some resonant mesonic
and baryonic states
which appear dynamically generated from the interactions and, thus, have a nature quite different to standard $q \bar{q}$ states. With some different formulations at the beginning \cite{juanenrique,Kaiser:1995cy,weise}, the more recent work uses the on shell formulation firstly established on the basis of the N/D method in \cite{nsd}, where the potential and the t-matrix in momentum space factorize outside the loop function implicit in the Bethe Salpeter equation in coupled channels that one uses in those approaches.  This is a very practical way to deal with the problem since one renders the coupled integral equations into a set of algebraic equations, paying a small price which is the fine tuning of some subtraction constant appearing in the dispersion relations involved. The approach is practical and useful, but carries also a handicap which is that one deals with amplitudes in momentum space, and couplings of the dynamically generated resonances to the different channels, and nowhere do wave functions in coordinate space appear in the approach. For instance, all properties of a resonance are given in terms of the mass and width and its couplings to the different channels, obtained from the residues of the amplitudes at the poles in the complex plane. Any intuition about the wave function of the different channels, its magnitude and extent in space, is lost in the approach as well as the meaning of the resonances and the discrete values obtained for the resonant energies.  Obviously, these are magnitudes that help to understand the microscopical composition of the dynamically generated states and, hence, a most welcome information.
However, this is not all, since the scattering amplitudes do not contain 
all the information of the wave function. They reflect the wave function 
at long distances. For some observables the wave function at small 
distances is needed. This is the case when one studies the response of 
states to external sources, which require to evaluate form factors, or 
expectation values of different observables. The wave functions, whether 
in momentum space or coordinate space, are then needed. To study 
couplings of the resonance to local sources the wave fucntion around the 
origin is required, but in the study of form factors one needs to know 
the wave function at all distances. We provide them both in the present 
work. They are already proving very useful to study new systems with 
three or more particles, when two of them cluster to form one of these 
dynamically generated resonances \cite{multirho}.
The first steps in this direction were done in \cite{conenrique} in order to understand the X(3872) resonance in terms of 
a molecule of $D^0 \bar{D^*}^0$ and $D^+ D^{*-}$ and their charge conjugates. 

We apply the formalism explained in Ref.~\cite{valencia2} to study the wave functions of the two $\Lambda(1405)$ states generated in the chiral unitary approach \cite{cola,carmina2,carmenjuan}.
For the description of the $\Lambda(1405)$ states, we use the chiral unitary approach~\cite{cola,oset}.
In this model there are two $\Lambda(1405)$ states dynamically generated in the coupled channels of meson-baryon scattering,~${\bar K}N,~\pi\Sigma,~\eta\Lambda$ and $K\Xi$.
The most interesting thing in this model is that two poles exist around the $\Lambda(1405)$ energy region at $z_1=(1390,-i66)$ MeV and $z_2=(1426,-i16)$ MeV \cite{cola}.
The electromagnetic mean squared radii of $\Lambda(1405)$ are calculated in Ref.~\cite{sekihara2} and are shown to be much larger than that of ground state baryons.
The $\Lambda(1670)$ was first reported as a dynamically generated resonance in Refs.~\cite{carmenjuan,oset} and has been corroborated in following works~\cite{carmina2,Hyodo:2006kg}.

\section{Wave functions}
We follow closely the formalism of \cite{conenrique} adapting it to the case of open channels.
The detail of our formalism is in Ref.~\cite{valencia2}.

Let us consider the Schr$\ddot{\rm o}$dinger equation in $N$ channel.
The first thing when dealing with coupled channels is that one must find the boundary conditions for the physical process that one is studying.
If we wish to create a resonance from the interaction of many channels at a certain energy we must take a channel which is open at this energy and make the two particles collide, starting from an infinite separation at $t=-\infty$.
Let us call channel 1 to this open channel that undergoes the scattering.
The equations to solve, with the appropriate boundary condition of a scattering state for channel 1 are
\begin{equation}
|\psi\rangle=|\phi\rangle+\frac{1}{E-H_0}V|\psi\rangle
\end{equation}
where
\begin{equation}
|\psi\rangle\equiv
\begin{Bmatrix}
|\psi_1\rangle\\
|\psi_2\rangle\\
\vdots \\
|\psi_N\rangle\\
\end{Bmatrix}
~~,|\phi\rangle\equiv
\begin{Bmatrix}
|\phi_1\rangle\\
0\\
\vdots \\
0\\
\end{Bmatrix}
\end{equation}
and $|\phi_1\rangle=|{\vec p}'\rangle$, such that ${\vec p}^{'2}/2\mu_1+M_1=E$, where we will use the notation $M_i=m_{1i}+m_{2i}$ and $\mu_i=m_{1i}m_{2i}/(m_{1i}+m_{2i})$.
$|\psi_i\rangle$ and $|\phi_1\rangle$ satisfy $(H_0+V)|\psi_i\rangle=E|\psi\rangle$ and $H_0|\phi_1\rangle=E|\phi\rangle$.

And we can write the wave function in coordinate space as

\begin{eqnarray}
\langle {\vec  x}|\psi_1\rangle&=&\frac{1}{(2\pi)^{3/2}}e^{i {\vec  p}\cdot{\vec  x}}+\int_{p<\Lambda}d^3p\frac{1}{(2\pi)^{3/2}}e^{i {\vec  p}\cdot{\vec  x}}\frac{1}{E-M_1-{\vec  p}^2/2\mu_1+i\epsilon}t_{11}(E)
\label{eq:xWF1}\\
\langle {\vec  x}|\psi_i\rangle&=&\int_{p<\Lambda}d^3p\frac{1}{(2\pi)^{3/2}}e^{i {\vec  p}\cdot{\vec  x}}\frac{1}{E-M_i-{\vec  p}^2/2\mu_i+i\epsilon}t_{i1}(E)~~(i\ne 1)
\label{eq:xWFi}
\end{eqnarray}
with $|{\vec p'}|=\sqrt{2\mu_1(E-M_1)}$.

\begin{figure}
  \includegraphics[height=.7\textheight]{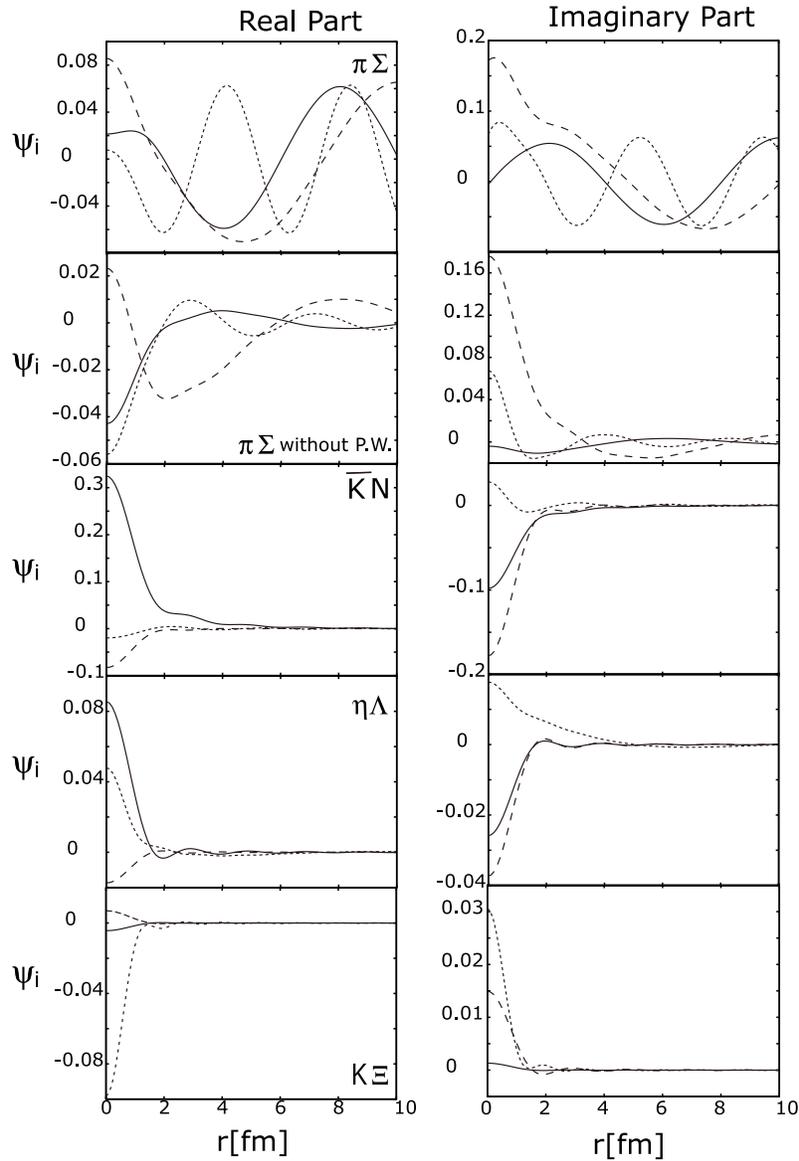}
\caption{\label{fig:WF}Wave functions in coordinate space. 
Solid lines and dashed lines show the results obtained at $E=1426$ MeV and at $E=1390$ MeV for the $\Lambda(1405)$ resonances, and dotted lines show that at $E=1680$ MeV for the $\Lambda(1670)$ state.
The figures of the first line contain the full wave function of $\pi\Sigma$ (initial plus scattered), while the second line shows only the scattered wave of the $\pi\Sigma$ state}

\end{figure}

We show in Fig.~\ref{fig:WF} the wave function in coordinate space at the pole energies of the two $\Lambda(1405)$, together with that of the $\Lambda(1670)$.
From Fig.~\ref{fig:WF}, we find that the ${\bar K}N$ components dominate at 1426 MeV while the $\pi\Sigma$ components are dominant at 1390 MeV.
This is consistent with the findings of \cite{cola,carmina2,carmenjuan, borasoy, ollerKa,borasoyulf} that the pole at higher energies couples most strongly to ${\bar K}N$ while the one at lower energies couples mostly to $\pi\Sigma$.

The curves correspond to these different energies, which are the energies where we find the poles of the $\Lambda(1390)$, $\Lambda(1426)$ and $\Lambda(1670)$.
We can see how the wave function concentrates close to the origin, and both for bound channels, as well as for open channels, fades away rapidly beyond 2 $fm$, providing a spatial picture of the distribution of the particles of the different channels building up the resonances.

In the case of the $\Lambda(1670)$ we can see in the last line of Fig.~\ref{fig:WF} that the dominant component is the $K\Xi$ bound state.
This resonance would very approximately qualify as a $K\Xi$ bound state, as also suggested in Ref.~\cite{oset} based on the large couplings of the resonance to that state.

\section{Summary}
    In this work we have developed a formalism to deal with coupled channels in a unitary approach by paying a special attention to constructing the wave functions in the different channels in the case that there are resonances dynamically generated. The paper generalizes what was found before for only bound coupled channels. Here we have bound and open channels and the formalism is subtly different, since contrary to the case of bound states, where only discrete energies are allowed, here we have a continuous energy variable. Many of the results obtained for bound states do not hold for the resonance states. One of the things we do is to identify the meaning of a resonance in the coupled channel approach, and it emerges as an approximate bound state of a coupled channel which can decay into the open ones.  The formalism developed is easy, practical and useful.  A separable potential in coordinate space is chosen which leads to an on shell factorization of the Bethe Salpeter equations (Lippmann Schwinger in the nonrelativistic form), which allows to convert the coupled channel integral equations into trivial algebraic equations. The wave functions in momentum space are then found as trivial analytic functions, from where the wave functions in coordinate space can be easily evaluated.  The couplings of the resonance to the different channels are related to the wave function at the origin and interesting relationships between these couplings are obtained.
     We also study the issue of couplings of the resonances to states outside the space of the building channels and justify results used before in the Literature, setting the limits for their application. Similarly, we also face the issue of final state interaction within the coupled channel formalism and find again a justification for results used in the Literature, setting again the limits of applicability. 
     
     As an application of the formalism, we tackle the problem of the two $\Lambda(1405)$ and the $\Lambda(1670)$ states dynamically generated in the chiral unitary approach from  the $\pi \Sigma$, $\bar{K} N$, $\eta \Lambda$, and $K \Xi$  interaction. We evaluate the wave functions in coordinate space for the first time, giving an intuitive idea of the wave functions and the spatial distribution of the particles of the different channels.  
     

\begin{theacknowledgments}
    This work is partly supported 
by DGICYT Contracts Nos. FIS2006-03438 and FEDER funds, FIS2008-01143 and CSD2007-00042 (CPAN), the Generalitat Valenciana in the program Prometeo 
and the EU Integrated
Infrastructure Initiative Hadron Physics Project under contract
RII3-CT-2004-506078.
This research is part of the European
 Community-Research Infrastructure Integrating Activity ``Study of
 Strongly Interacting Matter'' (acronym HadronPhysics2, Grant
 Agreement n. 227431) 
 under the Seventh Framework Programme of EU.  
\end{theacknowledgments}



\bibliographystyle{aipproc}   

\bibliography{sample}

\IfFileExists{\jobname.bbl}{}
 {\typeout{}
  \typeout{******************************************}
  \typeout{** Please run "bibtex \jobname" to optain}
  \typeout{** the bibliography and then re-run LaTeX}
  \typeout{** twice to fix the references!}
  \typeout{******************************************}
  \typeout{}
 }

\end{document}